\documentclass[fleqn,twoside,twocolumn,nofootinbib,showkeys]{revtex4} %
\usepackage[sec,nocpr,nopacs]{ujp} 

\begin{document}
\title[Autooscillatory Dynamics in a Mathematical Model of the Metabolic
Process]
{AUTOOSCILLATORY DYNAMICS\\ IN A MATHEMATICAL MODEL OF THE
METABOLIC\\ PROCESS IN AEROBIC BACTERIA. INFLUENCE\\ OF THE KREBS
CYCLE ON THE
SELF-ORGANIZATION\\ OF A BIOSYSTEM}%
\author{V.I.~Grytsay}
\affiliation{\bitp}
\address{\bitpaddr}
\email{vigrytsay@gmail.com}
\author{A.G.~Medentsev}%
\affiliation{G.K. Skryabin Institute of Biochemistry and Physiology of Microorganisms of the RAS}%
\address{5, Prosp. Nauki, Pushchino, Moscow region, Russian
Federation}%
\email{medentsev-ag@rambler.ru}
\author{A.Yu.~Arinbasarova\,}
\affiliation{G.K. Skryabin Institute of Biochemistry and Physiology of Microorganisms of the RAS}%
\address{5, Prosp. Nauki, Pushchino, Moscow region, Russian
Federation}%

\udk{577.3}  \razd{\seciii}

\autorcol{V.I.\hspace*{0.7mm}Grytsay, A.G.\hspace*{0.7mm}Medentsev,
A.Yu.\hspace*{0.7mm}Arinbasarova}

\setcounter{page}{1}%

\begin{abstract}
We have modeled  the metabolic process running in aerobic cells as
open nonlinear dissipative systems. The map of metabolic paths and
the general scheme of a dissipative system participating in the
transformation of steroids are constructed. We have studied the
influence of the Krebs cycle on the dynamics of the whole metabolic
process and constructed projections of the phase portrait in the
strange attractor mode. The total spectra of Lyapunov exponents,
divergences, Lyapunov's dimensions of the fractality,
Kolmogorov--Sinai entropies, and predictability horizons for the
given modes are calculated. We have determined the bifurcation
diagram presenting the dependence of the dynamics on a small
parameter, which defines system's physical state.
\end{abstract}
\maketitle

\section{Introduction}\vspace*{-1mm}

We will study the causes for the appearance of an autooscillatory
dynamics in the metabolic process running in aerobic bacteria which
was observed experimentally in a bioreactor with
\textit{Arthrobacter globiformis}  cells [1--5]. The consideration
is carried on with the help of the synergetic method of modeling of
the metabolic processes developed in works V.I.~Grytsay [6--33]. The
application of this method will allow us to consider the
self-organization and the dynamic chaos in the metabolic processes
in cells and in organism as a whole and to establish the physical
laws for their vital processes.

\begin{figure*}%
\vskip1mm
\includegraphics[width=15cm]{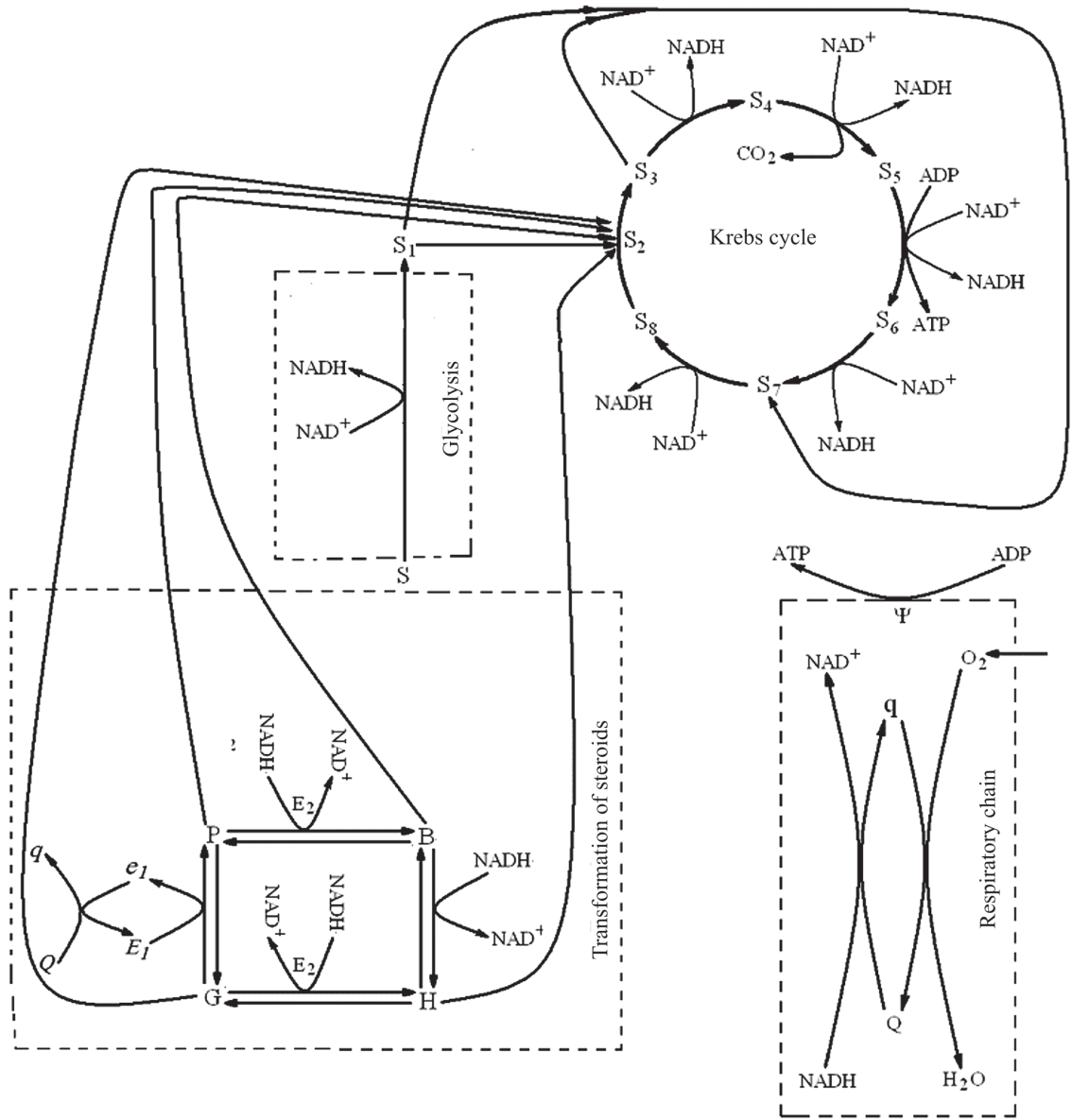}
\vskip-3mm\caption{Map of metabolic paths of a cell
\textit{Arthrobacter globiformis}  }
\end{figure*}

The mentioned microorganism is referred to oxy\-gen-breathing
bacteria arisen 2.48 bln years ago. Due to the evolution of the
metabolic processes of protobionts, the microorganisms not consuming
oxygen transited to oxygen-breathing bacteria with their subsequent
evolution to eukaryotes. The consideration of a specific biochemical
process of transformation of steroids allows us to use the
experimentally determined parameters in the construction of a model
and to conclude about structural-functional connections under
conditions of the self-organization in the vital activity of cells.
The relative simplicity of the given metabolic process of bacteria
under study gives possibility to model the whole metabolic process
in a cell as an open dissipative system, in which two main systems
(namely, the system of transformation of a substrate and the
respiratory chain) necessary for its life are self-organized.

\section{Mathematical Model}

In  order to develop a mathematical model of the process of
transformation of steroids by aerobic bacteria \textit{Arthrobacter
globiformis}, we use the map of their metabolic paths at the
transformation of steroids (Fig.~1) obtained as a result of the
processing of experimental data [2, 3].

Since we consider only the process of transformation of steroids, we
separate the localized dissipative system of transformation of steroids,
which interacts with other metabolic processes in a cell. Its general scheme
is presented in Fig. 2.

Based on the scheme in Fig. 2, we developed a mathematical model of the
metabolic process in a cell in the following form:
\begin{equation}
\label{eq1} \frac{dG}{dt}=\frac{G_0 }{N_3+G+\gamma_2\psi}-l_1
V(E_1)V(G)-\alpha_3G,
\end{equation}\vspace*{-7mm}
\begin{equation}
\label{eq2}\frac{dP}{dt}=l_1 V(E_1)V(G)-l_2 V(E_2 )V(N
)V(P)-\alpha_4P,
\end{equation}\vspace*{-7mm}
\begin{equation}
\label{eq3} \frac{dB }{dt}=l_2 V(E_2 )V(N)V(P)-k_1
V(\psi)V(B)-\alpha_5B,
\end{equation}\vspace*{-7mm}
\[
\frac{E_1
}{dt}=E_{10}\frac{G^2}{\beta_1+G^2}\left(\!1-\frac{P+mN}{N_1+P+mN}\!\right)-\]\vspace*{-7mm}
\begin{equation}
\label{eq4} -\, l_1 V(E_1)V(G)+l_4V(e_1)V(Q)-\alpha_1 E_1,
\end{equation}\vspace*{-7mm}
\begin{equation}
\label{eq5} \frac{de _1 }{dt}=-l_4 V(e_1
)V(Q)+l_1V(E_1)V(G)-\alpha_1e_1,
\end{equation}\vspace*{-7mm}
\[
\frac{dQ}{dt}=6lV(2-Q)V(O_2)V^{(1)}(\psi)\,-
\]\vspace*{-7mm}
\begin{equation}
\label{eq6} -\, l_6 V(e_1)V(Q)-l_7 V(Q)V(N),
\end{equation}\vspace*{-7mm}
\begin{equation}
\label{eq7}
\frac{dO_2}{dt}=\frac{O_{20}}{N_5+O_2}-lV(2-Q)V(O_2)V^{(1)}(\psi)-\alpha_7O_2,
\end{equation}\vspace*{-7mm}
\[
\frac{dE_2}{dt}=E_{20}\frac{P^2}{\beta_2+p^2}\frac{N}{\beta+N}\left(\!1-\frac{B}{N_2+B}\!\right)-
\]\vspace*{-7mm}
\begin{equation}
\label{eq8} -\,l_{10}V(E_2 )V(N )V(P)-\alpha_2E_2,
\end{equation}\vspace*{-7mm}
\[
\frac{dN}{dt}=-l_2 V(E_2 )V(N)V(P)-l_7 V(Q )V(N ))\,+
\]\vspace*{-7mm}
\begin{equation}
\label{eq9} +\,k_2 V(B
)\frac{\psi}{K_{10}+\psi}+\frac{N_0}{N_4+N}-\alpha_6N,
\end{equation}\vspace*{-7mm}
\begin{equation}
\label{eq10}\frac{d\psi}{dt}=l_5V(E_1)V(G)+l_8V(N)V(Q)-\alpha\psi,
\end{equation}
where $V(X)=X/(1+X)$; $V^{(1)}(\psi )=1/(1+\psi ^2)$; $V(X)$ is a
function describing the adsorption of the enzyme in the region of a
local bond; and $V^{(1)}(\psi )$ is a function characterizing the
influence of a kinetic membrane potential on the respiratory chain.

In the modeling, it is convenient to introduce the following
dimensionless parameters: $l=l_1 =k_1 =0.2$; $l_2 =l_{10} =0.27;$
$l_5 =0.6$; $l_4 =l_6 =0.5$; $l_7 =1.2$; $l_8 =2.4$; $k_2 =1.5$;
$E_{10} =3$; $\beta _1 =2$; $ N_1 =0.03$; $ m=2.5$; $ \alpha
=0.033$; $a_1 =0.007$; $ \alpha _1 =0.0068$; $E_{20} =1.2$; $ \beta
=0.01$; $ \beta _2 =1$; $ N_2 =0.03$; $ \alpha _2 =0.02$; $ G_0
=0.019$; $ N_3 =2$; $ \gamma _2 =0.2$; $ \alpha _5 =0.014$; $ \alpha
_3 =\alpha _4 =\alpha _6 =\alpha _7 =0.001$; $ O_{20} =0.015$; $ N_5
=0.1$; $ N_0 =0.003$; $N_4 =1$; $ K_{10} =0.7$.

\begin{figure}%
\vskip1mm
\includegraphics[width=6cm]{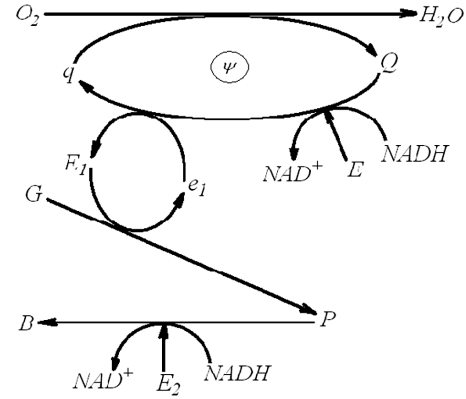}
\vskip-3mm\caption{General scheme of the process of transformation
of steroids by a cell \textit{Arthrobacter globiformis}  }
\end{figure}

\begin{figure}%
\vskip1mm
\includegraphics[width=\column]{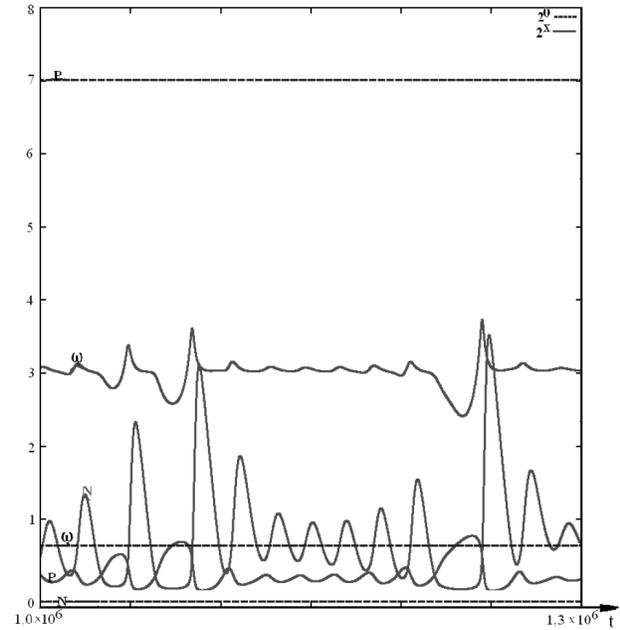}
\vskip-3mm\caption{Kinetic curves for some variables of the
stationary $2^0$ ($N_0 = 0$) and autooscillatory $2^n$ ($N_0 =
0.00366$) modes  }
\end{figure}

Equations (1)--(9) describe variations in the concentrations of
hydrocortisone ($G)$ (1); prednisolone ($P)$ (2); 20$\beta
$-oxyderivative of prednisolone ($B)$ (3); oxidized form of
3-ketosteroid-$\Delta '$-dehydrogenase ($E_1 )$ (4); reduced form of
3-ketosteroid-$\Delta '$-dehydrogenase ($e_1 )$ (5); oxidized form
of the respiratory chain ($Q)$ (6); oxygen ($O_2 )$ (7); $20\beta
$-oxysteroid-dehydrogenase ($E_2 )$ (8); and (9) $NAD\cdot H$(
reduced form of nicotinamide adenin dinucleotide) ($N)$. Equation
(10) shows a change in the kinetic membrane potential~($\psi )$.

\begin{figure*}%
\vskip1mm
\includegraphics[width=14.47cm]{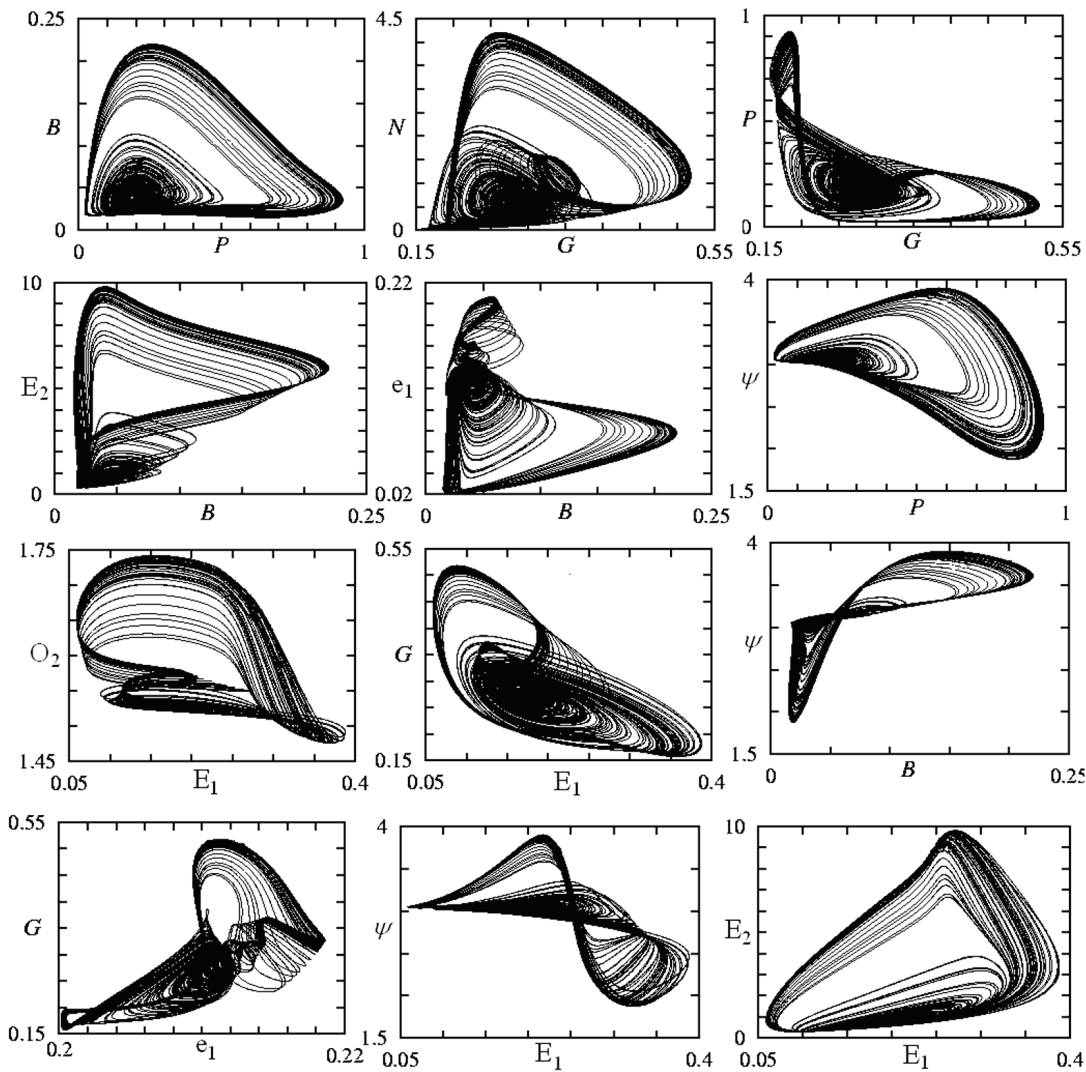}
\vskip-3mm\caption{Projections of the phase portrait of a strange
attractor arising at $N_0 = 0.00344$  }\vspace*{-2mm}
\end{figure*}

\begin{figure*}%
\vskip1mm
\includegraphics[width=14.5cm]{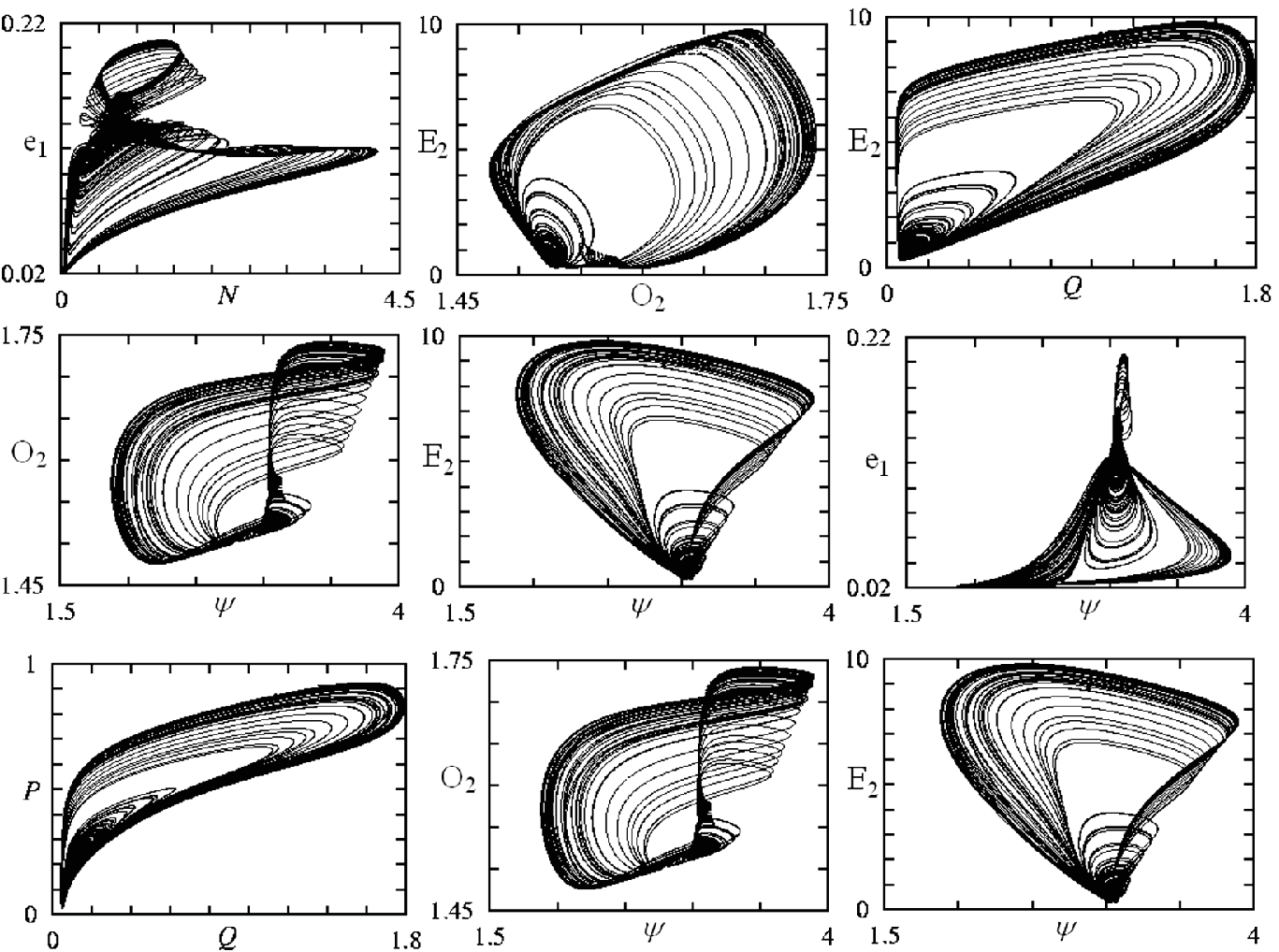}
\vskip-3mm\caption{Projections of the phase portrait of a strange
attractor arising at $N_0 = 0.00344$  }
\end{figure*}

\begin{figure*}%
\vskip1mm
\includegraphics[width=12cm]{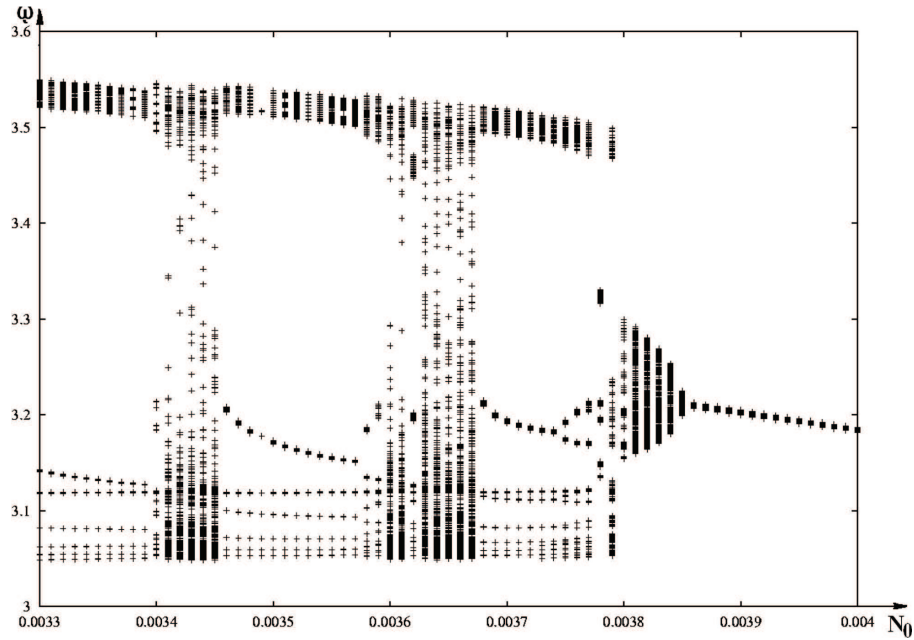}
\vskip-3mm\caption{Bifurcation diagram of the dependence of the form
of attractors of the dynamic process on the parameter $N_0 $ }
\end{figure*}

The reduction of parameters of the system to the dimensionless form was
presented in [2, 3].

The calculations according to the given mathematical model (1)--(10)
were carried out with the application of the theory of nonlinear
differential equations [34].

The analogous modeling of bioprocesses was realized in a lot of
works (see, e.g., [35--40]).

\section{Results of the Study}\vspace*{-1mm}

Within the constructed mathematical model (1)--(10), we performed
the computational experiments, by studying the dependence of the
kinetics of the metabolic process in a cell on the Krebs cycle [26].
These both processes are coupled with each other by the level of
NADH (N). The variation in the amount of this metabolite during the
Krebs cycle affects the respiratory chain and the activity of enzyme
$E_2 $ (see Fig. 2). In Fig. 3, we present the kinetics of some
components in two modes: for $N_0 =0$ and $N_0 = 0.0366$. The change
in this parameter causes the transition from the stationary mode
$2^0$ to the autooscillatory one $2^n$. Such modes were observed in
the experiment with a bioreactor [4, 5]. However, the nature of such
oscillations was not clarified, though some hypotheses were
advanced~[5].

\begin{table}[t]
\vskip4mm \noindent\caption{Total spectra of {Lyapunov exponents}
(\boldmath$\lambda _1 -\lambda _{10} )$, divergences ($\Lambda )$,
Kolmogorov--Sinai entropies ($h)$, foresight horizons ($t_{\min }
)$, and Lyapunov's dimensions of strange attractors ($D_{F_r } )$
for the dynamic modes at different $N_0 $}\vskip3mm\tabcolsep2.2pt

\noindent{\footnotesize\begin{tabular}{|c|l|l|l|l| }
 \hline
\raisebox{-11.5mm}[0cm][0cm]{\parbox[c][11mm][c]{10mm}{Lyapu-\\nov's\\
indices}} &
\multicolumn{4}{|c|}{\rule{0pt}{5mm}$N_0$}\\[2mm]%
 \cline{2-5}
\multicolumn{1}{|c|} {} & \multicolumn{1}{|c}{\rule{0pt}{5mm}0.003}&
\multicolumn{1}{|c}{0.00328720}& \multicolumn{1}{|c}{0.0043}&
\multicolumn{1}{|c|}{0.00344}\\[2mm] \cline{2-5}%
\multicolumn{1}{|c|} {} & \multicolumn{4}{|c|}{\rule{0pt}{5mm}Structure}\\[2mm] \cline{2-5}%
\multicolumn{1}{|c|} {} & \multicolumn{2}{|c}{\rule{0pt}{5mm}$2^n$}&
\multicolumn{2}{|c|}{$2^x$}\\[2mm]%
\hline%
\rule{0pt}{5mm}$\lambda _1 $&--0.0000056&~\,0.0000280&~\,0.0005342&~\,0.000607\\
$\lambda _2 $& --0.004731&--0.000586&~\,0.000019&~\,0.000019\\
$\lambda _3 $& --0.0049847&--0.0046497&--0.0047941&--0.00478731\\
$\lambda _4 $&--0.0083491&--0.0082351&--0.0079388&--0.00776693\\
$\lambda _5 $&--0.0230134&--0.0237059&--0.0230698&--0.023414386\\
$\lambda _6 $&--0.0302575&--0.0290098&--0.0292015&--0.0289861\\
$\lambda _7 $&--0.079155&--0.0796078&--0.80446039&--0.080484059\\
$\lambda _8 $&--0.0870153&--0.0860939&--0.0833784&--083210056101\\
$\lambda _9 $&--0.1807851&--1.7927866&--0.1787882&--0.178729\\
$\lambda _{10} $&--0.5214070&--0.5136290&--0.5147199&~\,0.5145829\\
$\Lambda $&--0.9354453&--0.9245716&--0.9217877&--0.921326\\
$h$&&&~\,0.000549&~\,0.000627\\
$t_{\min } $&&&~\,1821161&~\,1595914\\
$D_{F_r } $&&&~\,2.114537&~\,2.13088\\[2mm]
\hline
\end{tabular}
 }\vspace*{-2mm}
\end{table}

In the investigation of the physical dynamical pattern of the
mentioned oscillations in cells, we tested the observed
autooscillations for the stability by Lyapunov. For different values
of $N_0 $, we calculated the total spectra of {Lyapunov exponents}
(see Table 1), which enabled us to establish the dynamics of the
process. The form of constructed attractors characterizes the mode
of the self-organization of the metabolic process in a cell or the
mode of dynamic chaos as the transient mode describing the
adaptation of the metabolism to a change in the nutrition of a cell
from the external medium. By the determined values of {Lyapunov
exponents} for strange attractors, we calculated the Lyapunov
dimensions of their fractalities, Kolmogorov--Sinai entropies, and
foresight horizons [35]. On the basis of those data, we may judge
about the structure of strange attractors. Some their projections
are shown in Fig. 4,~5 ($N_0 = 0.00344$).

Then we calculated the bifurcation diagram presenting the dynamics
of the metabolic process as a function of $N_0 $ (see Fig. 6).
There, the transitions from the 1-fold mode to multiple modes, as
well as strange attractors, are clearly seen.\vspace*{-2mm}

\section{Conclusions}

A mathematical model of the open dissipative system with localized
metabolic process involving aerobic bacteria is presented. The
general map of its metabolic paths is constructed.

The synergetic method to study the self-organization and dynamical
chaos in metabolic processes in a cell and the whole organism is
developed.

In adreement with experimental data, we have determined the map of
paths of the metabolic process running in aerobic bacteria and the
general scheme of a dissipative system of transformation of
steroids. Using the constructed mathematical model, we have studied
the dependence of the dynamics on a change in the small parameter of
the Krebs cycle and found the modes of autooscillations and strange
attractors. The total spectra of {Lyapunov exponents}, divergences,
Lyapunov's dimensions of the fractality, Kolmogorov--Sinai
entropies, and predictability horizons are calculated. The
bifurcation diagram presenting the dependence of the dynamics of the
process on a small parameter determining the mode of
self-organization or dynamical chaos in the cell metabolism is
\mbox{constructed.}

The obtained scientific results have also practical meaning, by
presenting a physical interpretation of the causes for the
appearance of destructive autooscillatory modes observed in
biotechnological processes running in bioreactors (Section~2 and [4,
5]). The variables of the mathematical model which depend on a small
parameter will allow the bioengineers to competently control the
course of a biotechnological process.

\vskip3mm \textit{The present work was partially supported by the
Program of Fundamental Research of the Department of Physics and
Astronomy of the National Academy of Sciences of Ukraine
``Mathematics models of non-equilibrium process in open system''
No.~0120U100857.}


\rezume{%
В.Й.\,Грицай, О.Г.\,Меденцев, Г.Ю.\,Арінбасарова}{АВТОКОЛИВНА
ДИНАМІКА\\ В МЕТАБОЛІЧНОМУ ПРОЦЕСІ МАТЕМАТИЧНОЇ\\ МОДЕЛІ  АЕРОБНОЇ
БАКТЕРІЇ. ВПЛИВ ЦИКЛУ\\ КРЕБСА НА САМООРГАНІЗАЦІЮ БІОСИСТЕМИ}
{Проведено моделювання метаболічного процесу  аеробної клітини як
відкритої нелінійної дисипативної системи. Побудована карта її
метаболічних шляхів і загальна схема дисипативної системи, яка
приймає участь у трансформації стероїдів. Досліджено вплив циклу
Кребса на динаміку в цілому  метаболічного процесу, побудовано
проекції фазового портрету в режимі дивного атрактора. Розраховані
повні спектри показників Ляпунова, дивергенцій, ляпуновські
розмірності фрактальності, ентропії Колмогорова--Сіная та
горизонтипередбачування в даних режимах. Побудована біфуркаційна
діаграма залежності динаміки від малого параметра, що впливає  на
фізичний стан системи. }

\end{document}